# Simulation of the impact of people mobility, vaccination rate, and virus variants on the evolution of Covid–19 outbreak

Corrado Spinella[1] and Antonio Massimiliano Mio[1,*]

[1] Dipartimento di Scienze Fisiche e Tecnologie per la Materia, Consiglio Nazionale delle Ricerche, Piazzale Aldo Moro 7, I–00185 Rome (Italy)
corrado.spinella@cnr.it
antonio.mio@cnr.it

## Abstract

We have further extended our compartmental model describing the spread of the infection in Italy. The model is based on the assumption that the time evolution of all of the observable quantities (number of people still positive to the infection, hospitalized and fatalities cases, healed people, and total number of people that has contracted the infection) depend on average parameters, namely people diffusion coefficient, infection cross–section, and population density. The model provides precious information on the tight relationship between the variation of the reported infection cases and a well defined observable physical quantity: the average number of people that lie within the daily displacement area of any single person. The extension of the model now includes self–consistent evaluation of the reproduction index, effect of immunization due to vaccination, and potential impact of virus variants on the dynamical evolution of the outbreak. The model fits very well the epidemic data, and allows us to strictly relate the time evolution of the number of hospitalized case and fatalities to the change of people mobility, vaccination rate, and appearance of an initial concentration of people positives for new variants of the virus.

## Introduction

The new coronavirus severe acute respiratory syndrome coronavirus 2 (SARS–CoV–2), initially started in the city of Wuhan, China,[1–4] has transformed into a pandemic that has affected a large number of countries around the world.[5–7] Models are extremely useful to identify physical key parameters influencing the spread of infection and thus taking appropriate measures to limit serious consequences of the influenza/SARS pandemics.[3–4,7–14]

In this work we present a further extension of our theoretical description,[8] based on the assumption that spreading of viral infection can be described by a simple diffusion process, controlled mainly by a diffusion coefficient that changes in time accordingly to the people mobility restriction measures adopted in the course of the outbreak. The description is based on a compartmental model that allows us to accurately follow the time evolution of the observable quantities characterizing the virus outbreak: people tested positive for the virus, people tested as healed (i.e. negative, after a period from the infection, to the test for the virus), hospitalized people, fatalities, and total number of those who has been infected. We report on the tight correlation between people mobility and time evolution of hospitalized and fatalities cases, by examining, in particular, the data of the outbreak in Italy. The effect of vaccination is now included in the model, in order to analyze the best condition of easing the mobility restriction measures as a function of the implemented daily vaccination rate.

Most recently, there is also an increasing concern regarding to the possible appearance of virus variants, characterized by higher level of transmissibility or symptom severity. The model allows us to describe these effects and their impact on the risk of triggering new epidemic waves. Our approach allows us to get a fast feedback of the restriction measures adopted to reduce people mobility and a prediction of the evolution scenarios on the basis of a fit to the experimental available data.

## Experimental data: the Covid–19 outbreak in Italy

In Italy, as of January the 31$^{th}$, 2021, a total of 2,553,032 cases of coronavirus disease 2019 (COVID–19) and 88,516 deaths have been confirmed.[15] These data are shown in Fig. 1, in semi–logarithmic plots as a function of time. The data refer to the numbers of hospitalized people (open circles), people tested positive for the virus (open triangles), and fatalities (open squares) of the Covid–19 outbreak in Italy as a whole [Fig. 1(*a*)], and in three different National Italian Regions: Lombardia [Fig. 1(*b*)], Sicily [Fig. 1(*c*)], and Lazio [Fig. 1(*d*)], since February 2020, the 24$^{th}$. After the initial sudden increase of the number of cases, Italy implemented measures aimed to limit people mobility from March 2020 the 9$^{th}$ to June 2020 the 14$^{th}$. The consequence of such measures was a significant slowdown of the outbreak diffusion, following by a decrease in the number of positive and hospitalized cases, extended until end of July 2020. Easing of mobility restriction measures induced a new increase of cases, by triggering the second wave of the outbreak lasting until nowadays.

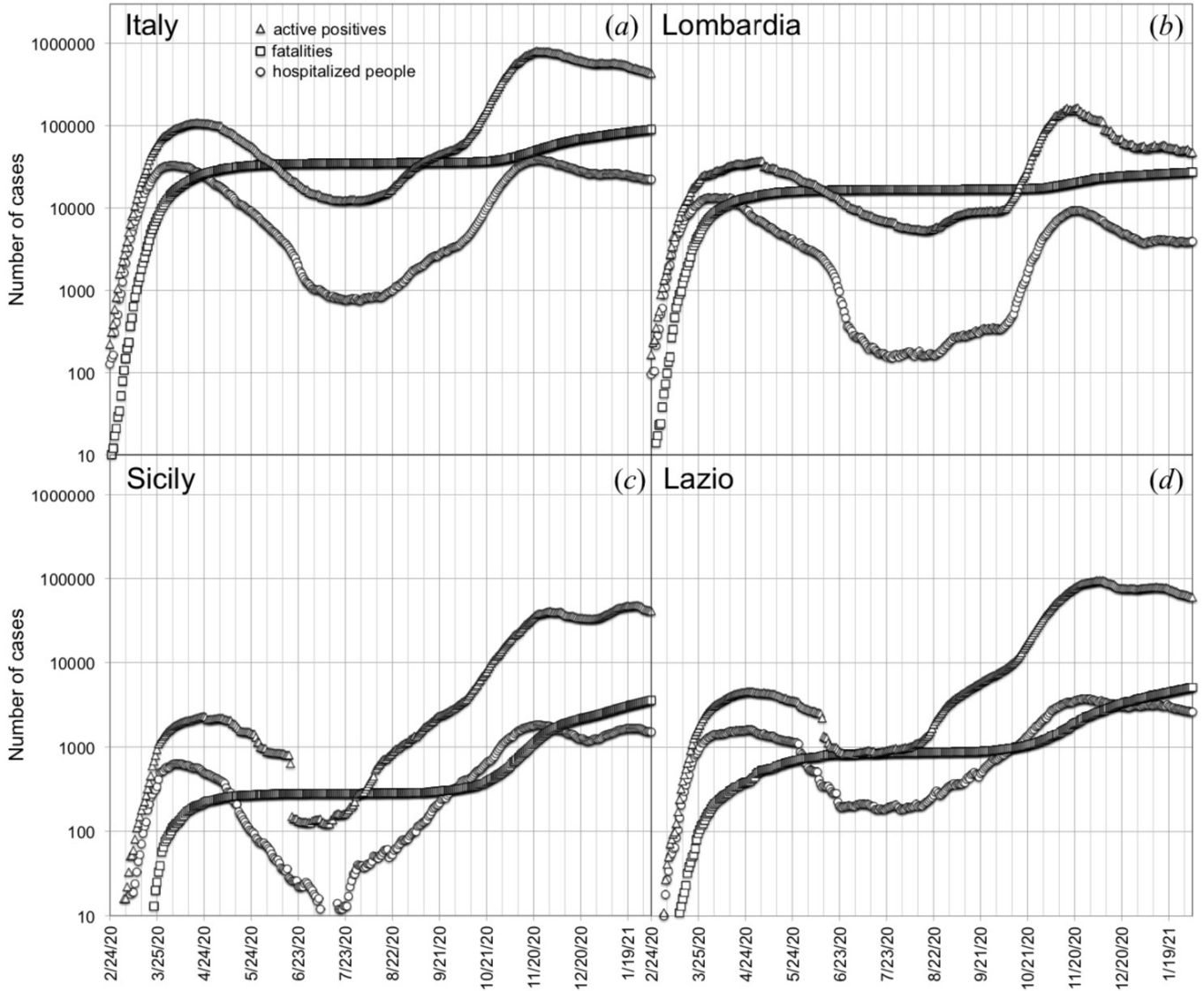

Fig. 1 Experimental data of hospitalized people (open circles), fatalities (open squares), people tested positive for the viral infection (open triangles), since the beginning of the Covid–19 outbreak in Italy (*a*), in Lombardia (*b*), in Sicily (*c*), and in Lazio (*d*).

## Methods

The model proposed to describe time evolution of the total number of infected people, positive cases, healed people, deaths, and hospitalized people, during the Covid–19 outbreak is based on a mean–field approximation. This consists in the assumption that the probability for an individual to contract the infection is uniformly proportional to the concentration $p$ of positive circulating cases, to a diffusion coefficient $D$, equal to the surface area covered on average by each person in a day, and to an infection cross–section $\sigma^2$ related to the probability of a single infection event ($\sigma^2 = \pi R^2$, $R$ being the average distance within which a healthy person can be infected by a positive one). This cross–section is a quantity specifically dependent on the virus infectiousness. We assume it is constant everywhere all over the examined geographic area and can change only in the presence of virus variants. In particular, the dimensionless probability $\eta$ of a single infection event is related to the infection cross–section by the relationship $\eta = \rho_0 \sigma^2$, where $\rho_0$ is the density of inhabitants. Under these hypotheses, at any instant $t$, the increase $dp$ of people positive for viral infection in the time interval $dt$ can be expressed as:

$$\frac{dp}{dt} = \rho_0 D \sigma^2 (\rho_0 - c)p - \frac{dg}{dt} - \frac{dm}{dt} \tag{1}$$

$g$ is the concentration of healed people (infected people who are tested negative for the virus after a certain time interval from the infection), $m$ is the concentration of fatalities, and $c$ is the total concentration of those who have contracted the virus at the time $t$.

A fraction $f$ of the new positive cases requires hospital care and, consequently, the concentration $r$ of hospitalized people will change, in the time interval $dt$, by a quantity $dr$ given by:

$$\frac{dr}{dt} = f \rho_0 D \sigma^2 (\rho_0 - c)p - \left(\frac{q}{\tau_1} + \frac{1-q}{\tau_2}\right) r \tag{2}$$

where we further consider that $r$ diminishes, in the same time interval $dt$, because a fraction $q$ of hospitalized people dies in a characteristic time $\tau_1$, whilst the complementary fraction $(1-q)$ heals in a characteristic time $\tau_2$. As a consequence, the concentration $m$ of fatalities will vary with time according to the following equation:

$$\frac{dm}{dt} = \frac{q}{\tau_1} r \tag{3}$$

While the fraction $f$ of positive people is hospitalized, the fraction $(1-f)$ does not exhibit serious symptoms until complete healing. Their concentration $s$ will vary with time according to the following relationship:

$$\frac{ds}{dt} = (1-f)\rho_0 D \sigma^2 (\rho_0 - c)p - \frac{s}{\tau_3} \tag{4}$$

$\tau_3$ being the characteristic time toward healing for these individuals. As reported in Ref. 16, this characteristic time is typically larger than $\tau_2$, i.e. the one used for describing time dependent healing of the most severe hospitalized cases (Eq. 2). As a consequence, the time dependence of the concentration $g$ of healed people changes with time according to:

$$\frac{dg}{dt} = \frac{s}{\tau_3} + \frac{(1-q)r}{\tau_2} \tag{5}$$

Finally, the total concentration $c$ of those who have contracted the infection will vary on time according to the following relationship:

$$\frac{dc}{dt} = \rho_0 D \sigma^2 (\rho_0 - c)p \tag{6}$$

Our description is based on the assumption that the dynamics of all the observable variables, $p$, $r$, $m$, $g$, $c$, can be described in terms of the time dependence of the diffusion coefficient $D = D$, while keeping constant the infection cross–section to the value of $= 3.14$ m$^2$ (corresponding to $R = 1$ m).

## Results and discussion

*Modelling the epidemic evolution in Italy before 2020 holiday season*

The values of the diffusion coefficient $D$, from February the 24$^{th}$, 2020 until December the 20$^{th}$, 2020 (i.e. a few days before holiday season in Italy), are plotted in Fig. 2(*b*) (open lozenges). These values were extracted from the data of the hospitalized cases [open circles in Fig. 2(*a*)] by adopting the analytical procedure described in detail in Ref. [8].

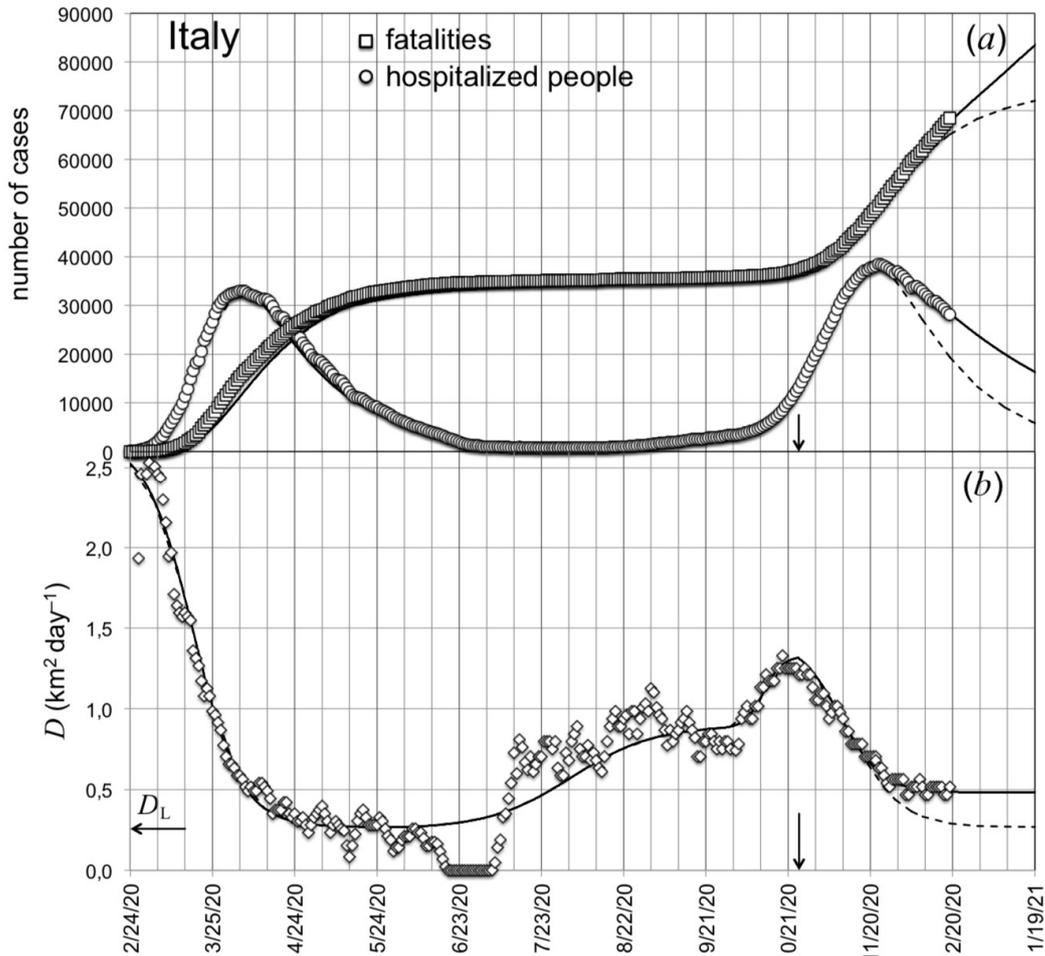

Fig. 2 (*a*) Evolution of the data of hospitalized people (open circles) and fatalities (open squares) in Italy during the Covid–19 outbreak until December the 20$^{th}$, 2020. (*b*) Corresponding values of the diffusion coefficient (open lozenges) extracted from the data of hospitalized cases. Continuous line in (*b*) is the result of a fit to the $D$ values by using a set of logistic functions (Eq. 7). This functional form is used to model hospitalized cases and fatalities represented by the continuous lines plotted in (*a*). Dashed lines represent the model simulation corresponding to a decrease of the diffusion coefficient, after October the 20$^{th}$, to the level, $D_L$, reached during the global spring lockdown.

We decided to extract the functional form of $D$ and all the other relevant model parameters from a fit to the data of hospitalized cases since they are more reliable than the ones concerning the number of people tested positive for the virus. Indeed, the latters represent only a small fraction of the real corresponding concentration values, since they refer to the cases actually detected through the adopted testing procedure (swabs), restricted to a defined relatively small sample of the entire population. In Fig. 2(*b*), the strong reduction of the diffusion coefficient to its minimum value $D_L$ =2.673×10$^5$ km$^2$ day$^{-1}$ is a consequence of the general lockdown in spring 2020, followed by a moderate increase during summer, when the mobility restriction rules were loosened. At the end of September 2020, people mobility increased at a higher rate, due to the resumption of school and work activity in more conventional modalities (compared to those based on work and school from home, experienced during the general spring lockdown). The fast increase of the diffusion coefficient in the first half of October 2020 has triggered the start of the second wave of the Covid–19 epidemic in Italy, accompanied by an exponential growth of the number of hospitalized people [Fig. 2(*a*)] in the following month.

On October the 25$^{th}$, 2020 [arrows in Fig. 2(*b*)] the Italian Government enacted further mobility restriction rules that induced a new decrease of the diffusion coefficient. These measures were significantly different Regions by Regions. A few of them, the so–called *red* Regions, experienced mobility restrictions very similar to the ones adopted during spring lockdown. For other Regions the measures were slightly less restrictive (*orange* Regions) up to a situation characterized by the persistence of a relatively high level of mobility with a limited number of restrictions (*yellow* Regions). The inhomogeneous intensity of the new mobility restriction rules reflects on the circumstance that the diffusion coefficient approached, at the beginning of December 2020, a constant value that was about 1.8 times higher than the one reached during the general, homogeneous spring lockdown. Continuous line in Fig. 2(*b*) is the result of a fit to the *D* values, as a function of time, by using a set of logistic functions of the kind of:[8]

$$D = D_2 + \frac{D_1 - D_2}{\exp\left(\frac{t - t_0}{\tau_c}\right) + 1} \qquad (7)$$

where $t_0$ is the time around which we observe change of the diffusion coefficient from $D_1$ to $D_2$ and $\tau_c$ is the characteristic duration of such a variation. In order to follow the variation of *D* as a function of time, the parameters $D_1, D_2, t_0, \tau_c$ where adjusted to their best fit values within four different time ranges: *i*) from February the 24$^{th}$ to June the 4$^{th}$, 2020 (start of the Covid–19 epidemic monitoring in Italy); *ii*) from June the 4$^{th}$ to September the 22$^{nd}$, 2020; *iii*) from September the 22$^{nd}$ to October the 25$^{th}$, 2020; *iv*) for times beyond October the 25$^{th}$, 2020. In particular, the last change of mobility (beyond October the 25$^{th}$, 2020) was modeled by setting $D_1$ =1.428×10$^6$ km$^2$day$^{-1}$, $D_2$ =4.811×10$^5$ km$^2$day$^{-1}$ (i.e., $D_2 = 1.8\, D_L$), $t_0 = 260$ days away from the start of the Covid–19 epidemic monitoring in Italy, and $\tau_c = 7.6$ days.

We used the functional form describing the variation of *D* as a function of time for calculating the expected values *r* (hospitalized people) and *m* (total number of fatalities), through Eqs. 1 to 6, and by adjusting the other model parameters by a fit to the data of Fig. 2(*a*), with the exception of the characteristic times $\tau_2$ (healing of hospitalized people) and $\tau_3$ (healing of infected, but not hospitalized people) that were set to the values found in the literature ($\tau_2 = 20$ days, $\tau_3 = 14$ days).[16] The results of such a procedure are the continuous lines plotted in Fig. 2(*a*). The agreement of the theoretical curves with the experimental data is excellent. The best–fit values found for *f* (fraction of the new infected persons that require hospitalization), $\tau_1$ (characteristic time for death), *q* (fraction of hospitalized people that die in the characteristic time $\tau_1$) were: *f* = 0.35%, $\tau_1$ = 7.2 days, *q* = 14% in the time range February 24$^{th}$ ≤ *t* ≤ May the 13$^{rd}$, *q* = 10% in the time range May 13$^{rd}$ < *t* ≤ November the 14$^{th}$, *q* = 15% for *t* > November the 14$^{th}$. In the same plots, dashed lines is the simulation of what would have occurred, according to the proposed theoretical model, if the diffusion coefficient, after October the 20$^{th}$, had approached the same value experienced in the occasion of the spring general lockdown.

The same analytical procedure was applied to model the evolution of epidemic data at a more geographically circumscribed level, by investigating the behavior of three Italian Regions that, on November the 5$^{th}$, 2020, were subjected to different restriction mobility measures: Lombardia ("red zone": severe mobility restriction measures), Sicilia ("orange zone": medium level of mobility restriction measures), Lazio ("yellow zone": extremely soft mobility restriction measures). The results are shown in Fig. 3. The model fits very well the hospitalized cases [full lines in Fig. 3(*a*),(*b*),(*c*)] and number of fatalities [full lines in Fig. 3(*d*),(*e*),(*f*)] by using the functional time dependences of *D* plotted as continuous lines in Fig. 3(*g*),(*h*),(*i*). Dashed lines are simulations of the behaviour we would have observed if the diffusion coefficient had decreased, after October the 20$^{th}$, 2020, to the corresponding spring lockdown values $D_L$. We notice that, in the range October the 20$^{th}$ – December the 20$^{th}$, the diffusion coefficient decreases to a plateau level that is different for the three examined Regions. Compared to the corresponding spring lockdown values $D_L$, the ratio $D/D_L$ approaches 1.5 for Lombardia, 1.8 for Sicily, and 2.2 for Lazio, respectively. Thus, the model allows us to strictly relate the temporal evolution of the epidemic data to the change of people mobility (diffusion coefficient) occurring as a consequence of restriction measures of different level of severity.

|  | *f* | $\tau_1$ (days) | *q* | | |
|---|---|---|---|---|---|
| Lombardia | 0.32% | 5.0 | 16%<br>Feb. the 24$^{th}$ ≤ *t* < Apr. the 18$^{th}$ | 6.5%<br>Apr. the 18$^{th}$ ≤ *t* < Nov. the 12$^{th}$ | 9%<br>*t* ≥ Nov. the 12$^{th}$ |
| Sicily | 0.30% | 5.1 | 6%<br>Feb. the 24$^{th}$ ≤ *t* < Apr. the 18$^{th}$ | 2.5%<br>Apr. the 18$^{th}$ ≤ *t* < Oct. the 25$^{th}$ | 11%<br>*t* ≥ Oct. the 25$^{th}$ |
| Lazio | 0.28% | 6.8 | 6%<br>Feb. the 24$^{th}$ ≤ *t* < May the 28$^{th}$ | 2.5%<br>May the 28$^{th}$ ≤ *t* < Oct. the 22$^{nd}$ | 8.6%<br>*t* ≥ Oct. the 22$^{nd}$ |

Tab. I  Parameter values used to fit the theoretical model to the data shown in Fig. 3

The model fitting parameters are listed in Tab. I. Also in this case (as for Italy as whole) we need to use values of the parameter *q* that are different within three different time range (the first range is centered on the first wave of the outbreak, the second one corresponds to the summer characterized by a relatively small number of cases, and the third range is around the peak of the second wave).

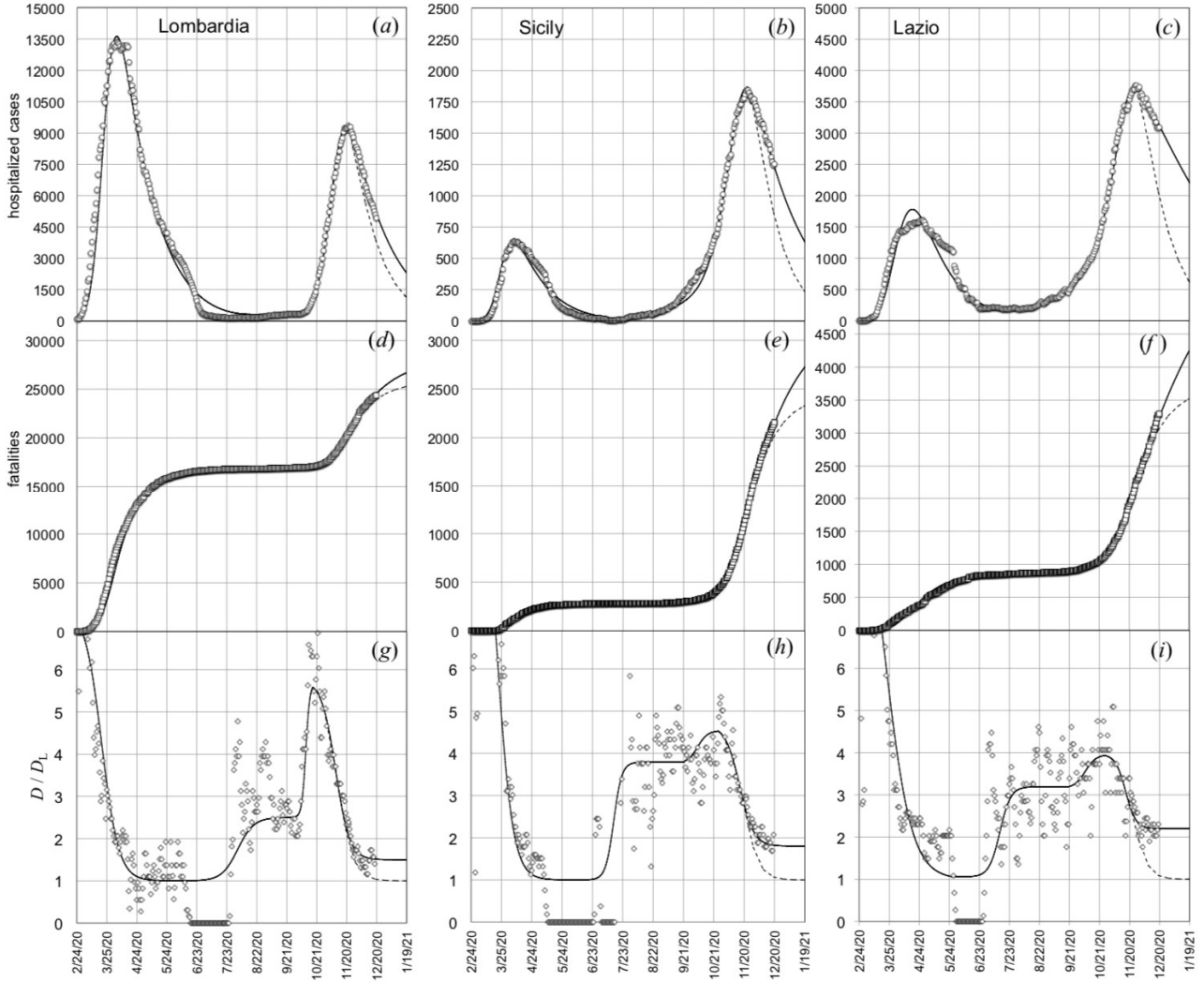

Fig. 3 (*a*), (*b*), (*c*) Hospitalized cases before December the 20$^{th}$, 2020, in Lombardia, Sicilia, and Lazio, respectively. (*d*), (*e*), (*f*) Corresponding number of fatalities. (*g*), (*h*), (*i*) Values of the diffusion coefficient normalized to the ones reached during the first lockdown in spring 2020, for Lombardia, Sicily, and Lazio, respectively. Continuous lines are fit to the data by using the model described in the text. Dashed lines are simulations of the behaviour we would have observed if the diffusion coefficient had decreased to $D_L$ (the spring lockdown value) after October the 20$^{th}$, 2020.

*Effect of mobility increase occurred during 2020 holiday season*

After December the 20$^{th}$, 2020, the Italian Government decided to slightly relax the mobility restriction measures in the occasion of holiday season. The corresponding increase of people mobility reflected in a significant slowdown of the decreasing rate of the hospitalized cases as shown in Fig. 4(*a*).

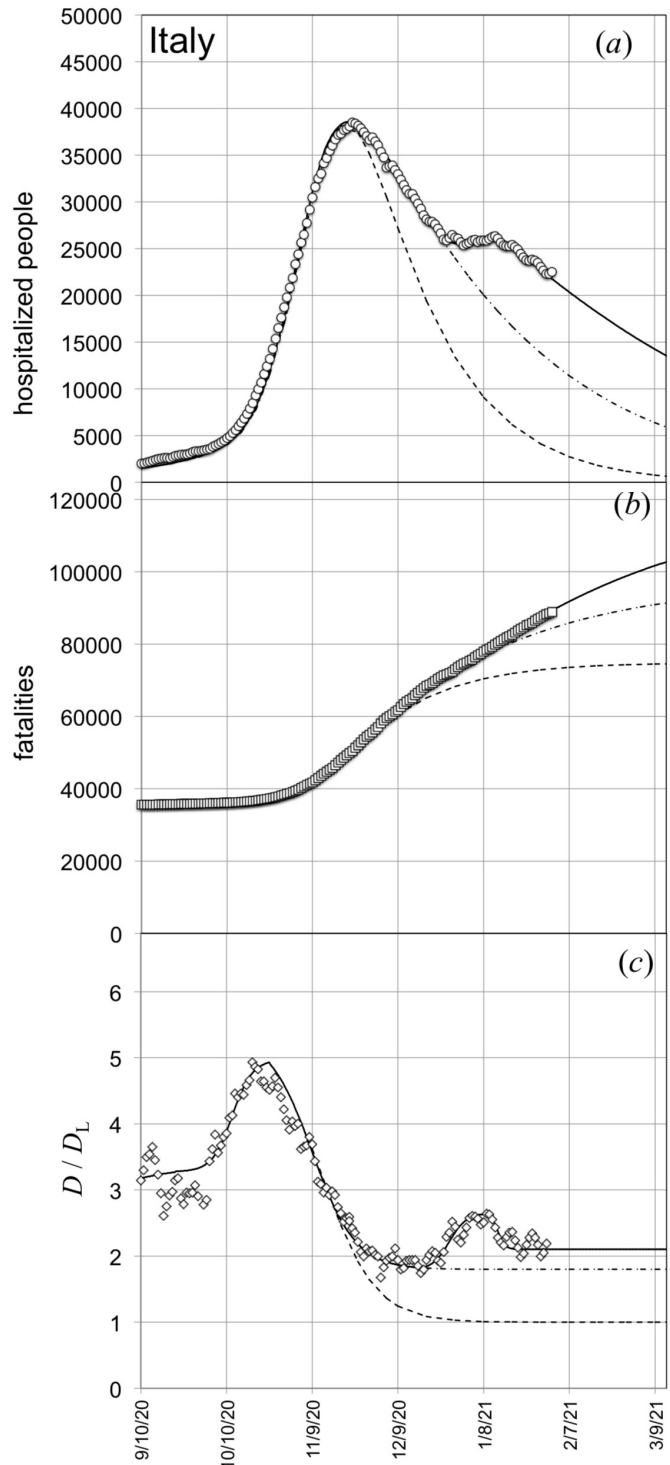

Fig. 4 (*a*) Hospitalized people and (*b*) fatalities in Italy in the time range centered on the second wave of the outbreak and on holiday season 2020. (*c*) Evolution of the diffusion coefficient, in the same time range, normalized to the spring lockdown value $D_L$. Continuous lines are fit of the model to the data. Dashed lines are reference curves corresponding to the hypothesis of a general lockdown on October the 20[th], 2020. Dot–dashed lines describe the situation we would have experienced if restriction mobility measures had maintained unchanged during holiday season.

The proposed theoretical description allows us to interpret this effect in terms of variation of the diffusion coefficient, referred again to the spring lockdown value $D_L$. In particular, Fig. 4(*c*) shows that the diffusion coefficient (and the related quantity $\rho_0 D$, i.e. "average number of people encountered by each person in a day") increased, during holiday season, from 1.8 to 2.6 times $D_L$. The reintroduction of more severe mobility restriction measures on January the 7$^{th}$ produced a new decrease of $D$. However, these new measures (mostly homogeneous in all the National territory) were significantly softer than the ones adopted before December the 20$^{th}$. This circumstance reflects on the observation that the ratio $D/D_L$ approached, after the holiday season peak, a value, for Italy as a whole, of about 2.1, appreciably higher than the one observed in the pre–peak time interval ($D/D_L$ = 1.8). The corresponding fits of the model to the hospitalized and fatalities cases are plotted as continuous lines in Fig. 4.

The simulation of the scenario expected in the presence of a decrease of the diffusion coefficient, after October the 20$^{th}$, 2020, to the same value of the spring lockdown is also plotted in Fig. 4 (dashed lines), whilst dot–dashed lines in the same figure refer to the scenario simulated by imposing that the diffusion coefficient had remained constant to the value experienced before holiday season.

The variation of $D/D_L$ around the peak value of holiday season has not been uniform among the different Italian Regions. This is clearly shown in Fig. 5 for Lombardia, Sicily, and Lazio. We notice that only Lazio has returned to a situation that evolves with a diffusion coefficient that is constant to the pre–holiday season value. Conversely, the ratio $D/D_L$ in Lombardia is approaching the value of 2.3, significantly larger than the pre–holiday season value ($D/D_L$ = 1.5). Sicily has experienced the most critical situation, with $D/D_L$ that as reached, during holiday season, a peak value as large as 3.6, thus triggering the start of a third wave of the outbreak, well visible in the plot of hospitalized cases shown in Fig. 5(*b*). Indeed, beyond the holiday season peak, the ratio $D/D_L$ for Sicily has exhibited a relatively slow decreasing rate and, on January the 15$^{th}$ [see the arrow in Fig. 5(*h*)], it was still at a level of about 3.3. In particular, dotted lines in Figs. 5(*b*),5(*e*),5(*h*) show the results of our model simulation under the hypothesis that $D/D_L$ for Sicily had approached, after the holiday season peak, a constant value equal to 3, demonstrating that this value would have been suitable to fuel the growth of a new wave of the virus epidemic in the Region. On January the 15$^{th}$, however, the Italian Government imposed for Sicily more severe mobility restriction measures ("red zone"), inducing a further decrease of $D/D_L$ and a consequent new diminution of the number of hospitalized people [Fig. 5(*b*)].

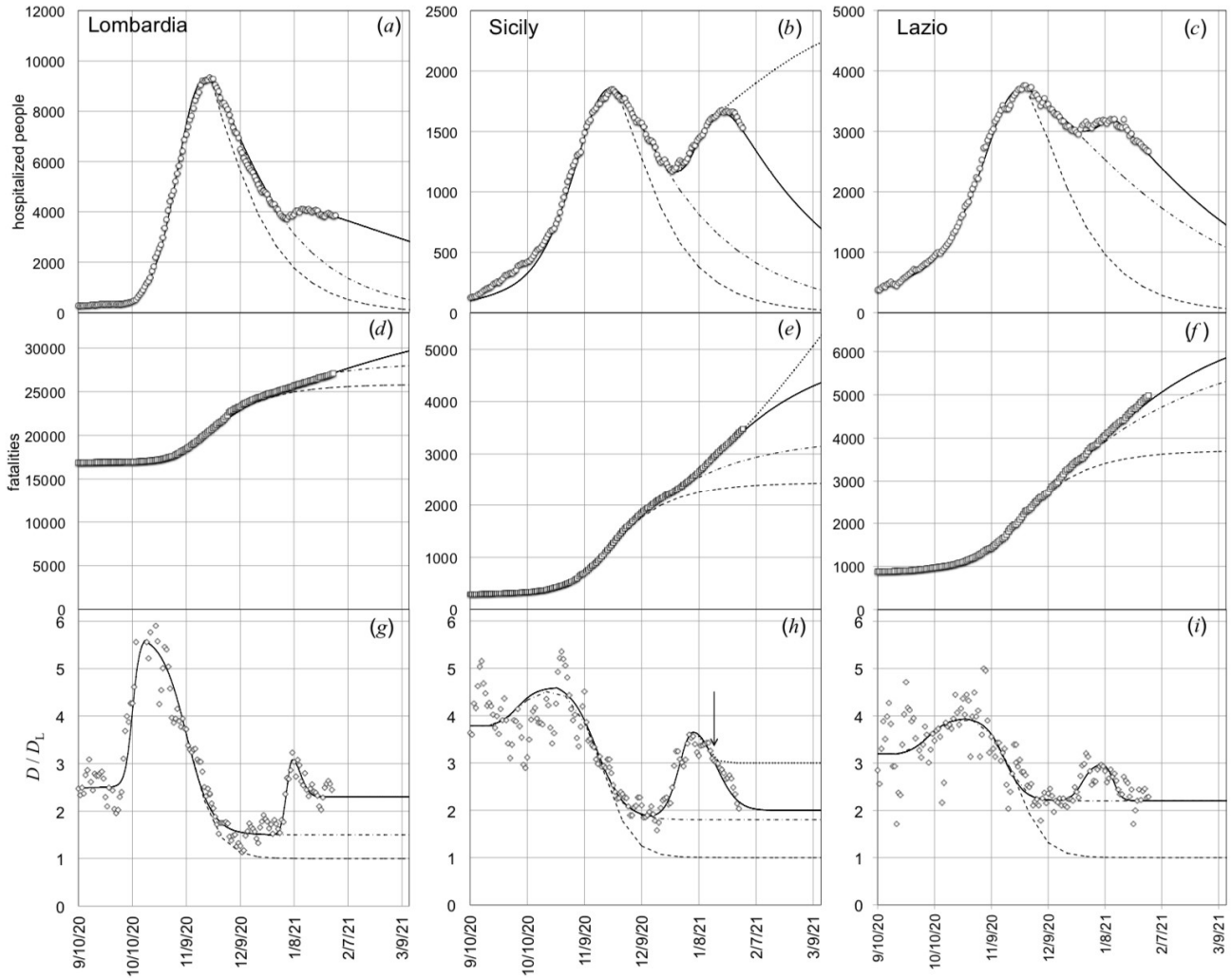

Fig. 5  (*a*), (*b*), (*c*) Hospitalized cases in the time range that includes holyday season 2020, in Lombardia, Sicilia, and Lazio, respectively. (*d*), (*e*), (*f*) Corresponding number of fatalities. (*g*), (*h*), (*i*) Values of the diffusion coefficient normalized to the corresponding spring 2020 values. Easing of restriction mobility measures during holiday season produced an increase of the diffusion coefficient, with peaks centered on January the 5$^{th}$, 2021. Continuous lines are fits of the model to the data. Dashed lines are reference curves corresponding to the decrease of *D*, after October the 20$^{th}$, 2020, to the same value of spring lockdown. Dot–dashed lines describe the situation we would have expected if restriction mobility measures had maintained unchanged during holiday season. For Sicily, the simulation of a post–peak holiday season diffusion coefficient that decreases to a level as large as 3 times than $D_L$ is shown (dotted lines).

*Modelling the impact of vaccine immunization*

On December the 27$^{th}$, 2020, Italy started its vaccination campaign. The investigation of the clinic effectiveness of the various vaccines that are being used is beyond the aim of this work. Our model, however, can include the influence of vaccination on the time evolution of the virus epidemic by assuming that immunization occurs, in general, about one week later from the inoculation of the second vaccine dose. Then, the concentration $\rho_i$ of people immunized by vaccination will increase with time *t* according to the following relationship:

$$\rho_i(t) = \int_{t_0+\tau}^{t} v(t-\tau)dt \tag{8}$$

where $t_0$ is the immunization time onset, corresponding to the day first person has received the second vaccine dose (January the 16$^{th}$, 2021, in our case), $\tau = 7$ days is the time interval for getting immunization from the second dose inoculation, and $v(t-\tau)$ is the number of people that was vaccinated (second dose) on the day corresponding to $t-\tau$.

Under these hypotheses, we can describe the influence of vaccine on the time evolution of all the observable variables, simply by substituting the term $(\rho_0 - c)$ with $(\rho_0 - \rho_i - c)$ in Eqs. 1 to 6. For the daily number of people, $v$, receiving the second vaccine dose, we used the data communicated by the Italian Health Ministry[17]. In order to simulate the impact of vaccination on the future time evolution of the virus outbreak, the function $v$ was kept constant to the vaccination daily rate experienced the week prior to the date of the last available data, (January the 31$^{st}$, 2021).

Vaccination is believed to be one of the best weapons we can use to strike virus outbreak and come back to highest levels of mobility in a relatively short time range. In order to investigate how vaccination can help us to increase people mobility, we have simulated several scenarios, shown in Fig. 6, consisting in a progressive increase of the diffusion coefficient to a level as high as the one reached at the end of summer 2020 ($D/D_L = 3.25$).

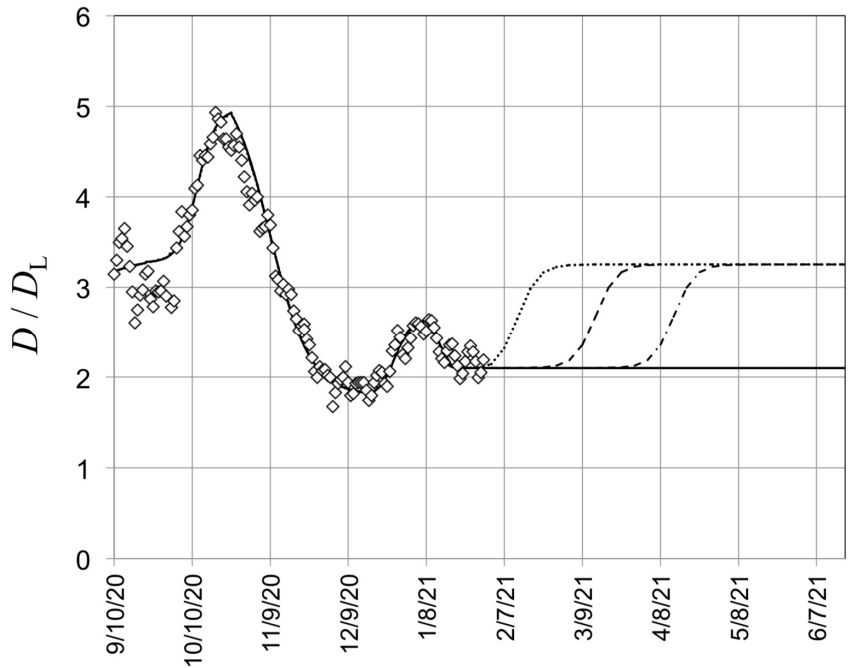

Fig. 6 Diffusion coefficient used for modelling people mobility during the second wave (in course) of the Covid–19 outbreak in Italy. The values are normalized to $D_L$, i.e. the level reached during the general lockdown of spring 2020. Dotted, dashed, and dot–dashed lines represent different possible scenarios of easing people mobility restriction measures to allow an increase of the diffusion coefficient to the level experienced at the end of summer 2020 ($D/D_L = 3.25$). Continuous line describes the situation according to which the diffusion coefficient is maintained constant to the value of January the 31$^{th}$, 2021 ($D/D_L = 2.1$)

These scenarios differ for the length of the time interval used for allowing the increase of the diffusion coefficient to the level of the end of summer 2020, starting from January the 31$^{st}$, 2021: 1 month (dotted line in Fig. 6), 2 months (dashed line in Fig. 6), and 3 months (dot–dashed line in Fig. 6). Instead, continuous line in Fig. 6 describes the situation of maintaining the diffusion coefficient constant to the present value ($D/D_L = 2.1$).

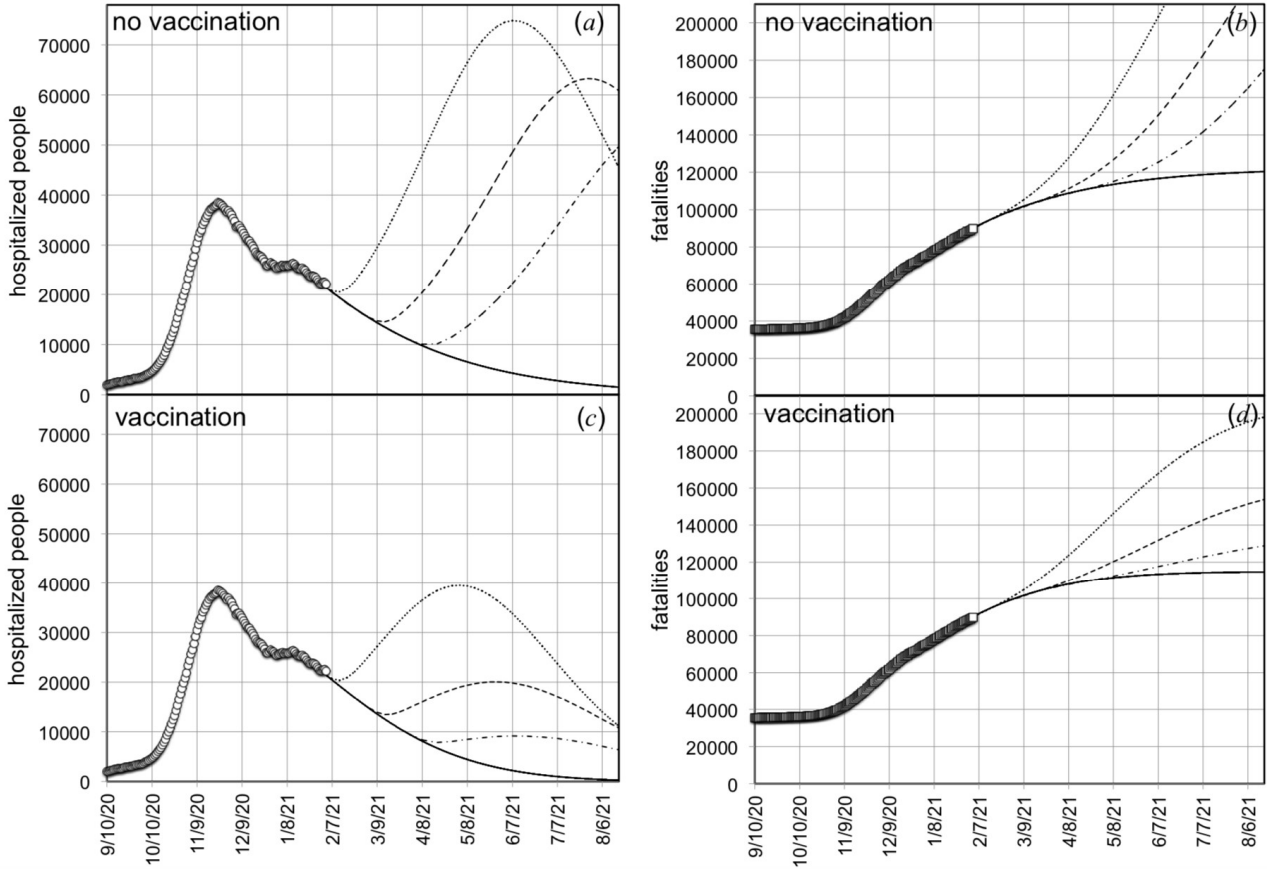

Fig. 7  Simulation of the number of hospitalized people and fatalities consequent to the increase of people mobility at the level experienced at the end of summer 2020, according to the time evolution of the diffusion coefficient shown in Fig. 6. Calculations in (*a*) and (*b*) were performed in the absence of vaccination, whilst in (*c*) and (*d*) we included the effect of vaccine immunization by assuming that the average vaccination daily rate keeps constant to the value of the last week, prior to January the 31$^{st}$, 2021.

The consequent model simulations of the time evolution of the number of hospitalized people and fatalities are plotted in Fig. 7, in the absence [Figs. 7(*a*),7(*b*)] or in the presence of vaccination [Figs. 7(*c*),7(*d*)]. We notice that vaccine immunization produces only negligible effects on the decreasing rate of hospitalized people and fatalities for the scenario consisting in maintaining constant the diffusion coefficient to the present level ($D/D_L = 2.1$), whilst it strongly mitigates the amplitude of the third wave peak of the outbreak triggered by the increase of people mobility to the levels measured at the end of summer 2020.

*Reproduction number calculation*

The proposed model provides a self–consistent method to evaluate the reproduction number $R_T$, i.e. the number of primary infection produced by a single infected person during the time interval he remains still positive for the virus (we actually assume that being positive for the virus is a suitable condition to transmit the infection with a constant probability, independently of the symptoms severity). In order to get such a result, we start by considering that the concentration $n$ of people infected by a concentration "probe" $\bar{p}(\theta)$, initially equal to $\bar{p}_t$, the new positives for the virus at the time instant $t$, normalized to $\bar{p}_t$ itself, increases, at a time instant $\theta \geq t$, by a quantity $dn$ given by:

$$\frac{dn}{\bar{p}_t d\theta} = \rho_0 D \sigma^2 [\rho_0 - \rho_i(\theta) - c(\theta)] \frac{\bar{p}(\theta)}{\bar{p}_t} \tag{9}$$

The function $c(\theta)$ is determined, for assigned forms of $\rho_i(\theta)$ and $D(\theta)$, by solving Eqs. 1 to 6, whilst $P(\theta) = \bar{p}(\theta)/\bar{p}_t$ decays with time according to the rate equations that describe the process of healing or death of people positives for the virus, i.e.:

$$\frac{dP}{d\theta} = -\frac{dG}{d\theta} - \frac{dM}{d\theta} \tag{10}$$

$$\frac{dR}{d\theta} = -\left(\frac{q}{\tau_1} + \frac{1-q}{\tau_2}\right) R(\theta) \tag{11}$$

$$\frac{dM}{d\theta} = \frac{q}{\tau_1} R(\theta) \tag{12}$$

$$\frac{dS}{d\theta} = -\frac{S(\theta)}{\tau_3} \tag{13}$$

$$\frac{dG}{d\theta} = \frac{S(\theta)}{\tau_3} + \frac{(1-q)R(\theta)}{\tau_2} \tag{14}$$

The analytical solution of this system of differential equations is straightforward:

$$P(\theta) = (1-f) \exp\left(-\frac{\theta}{\tau_3}\right) + f \exp\left[-\frac{\tau_1 + q(\tau_2 - \tau_1)}{\tau_1 \tau_2} \theta\right] \tag{15}$$

The function $P(\theta)$, calculated by setting $f, q, \tau_1, \tau_2, \tau_3$ to the values used to fit the model to the data of hospitalized people and fatalities for Italy as a whole, is plotted in Fig. 8. $P(\theta)$ is nothing but the probability that a single infected individual, contracting the viral infection at a given instant $\theta = 0$, is still positive and able, in turn, to infect susceptible people at a subsequent time $\theta > 0$.

$R_T$ is defined as the number of people infected by a single individual, positive for the virus at the certain instant $t$, throughout his full lifetime (until healing or death), and then:

$$R_T = \int_t^{+\infty} \frac{dn}{\bar{p}_t d\theta} d\theta = \rho_0 \sigma^2 \int_t^{+\infty} D(\theta)[\rho_0 - \rho_i(\theta) - c(\theta)]P(\theta)d\theta \qquad (16)$$

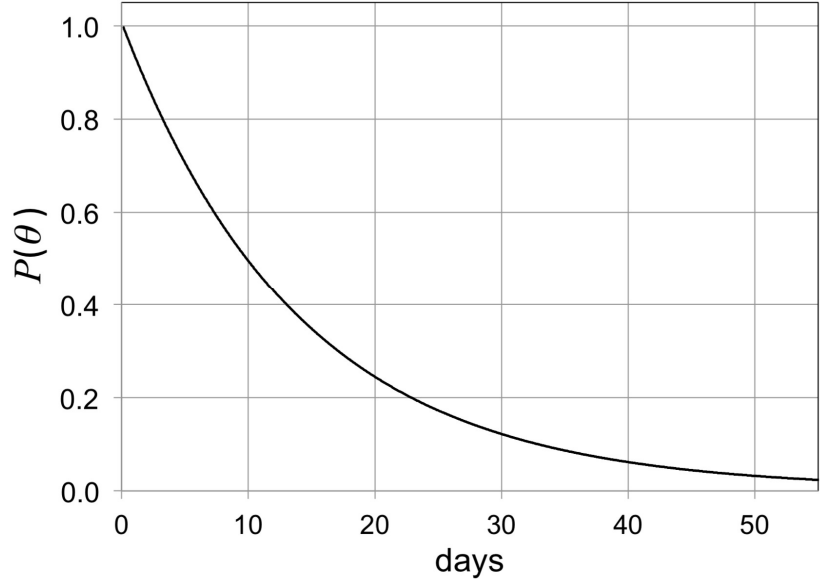

Fig. 8　Calculated probability that a single infected individual is still positive for the virus (and then able to infect, in turn, susceptible people) days after the initial instant of his infection.

Since $R_T$ is the result of a time integration, the instant, $T$, the determination of $R_T$ should be referred to is equal to the time average weighted on $dn$, i.e. the number of people that a single positive infects per unit of time throughout his lifetime:

$$T = \frac{\int_t^{+\infty} \theta \frac{dn}{d\theta} d\theta}{\int_t^{+\infty} \frac{dn}{d\theta} d\theta} = \frac{\int_t^{+\infty} \theta D(\theta)[\rho_0 - \rho_i(\theta) - c(\theta)]P(\theta)d\theta}{\int_t^{+\infty} D(\theta)[\rho_0 - \rho_i(\theta) - c(\theta)]P(\theta)d\theta} \qquad (17)$$

The time dependence of $R_T$ calculated by Eq. 16, as a function of the corresponding time $T$ expressed by Eq. 17, is plotted in Fig. 9 for the different scenarios of people mobility variations described in Fig. 6. By the direct comparison of Fig. 9 to Figs. 6 and 7, we notice that the increase of the diffusion coefficient to the level experienced at the end of summer 2020 in Italy, corresponds to the increase of the reproduction number above 1 and it is responsible for the triggering of a new wave of the virus outbreak.

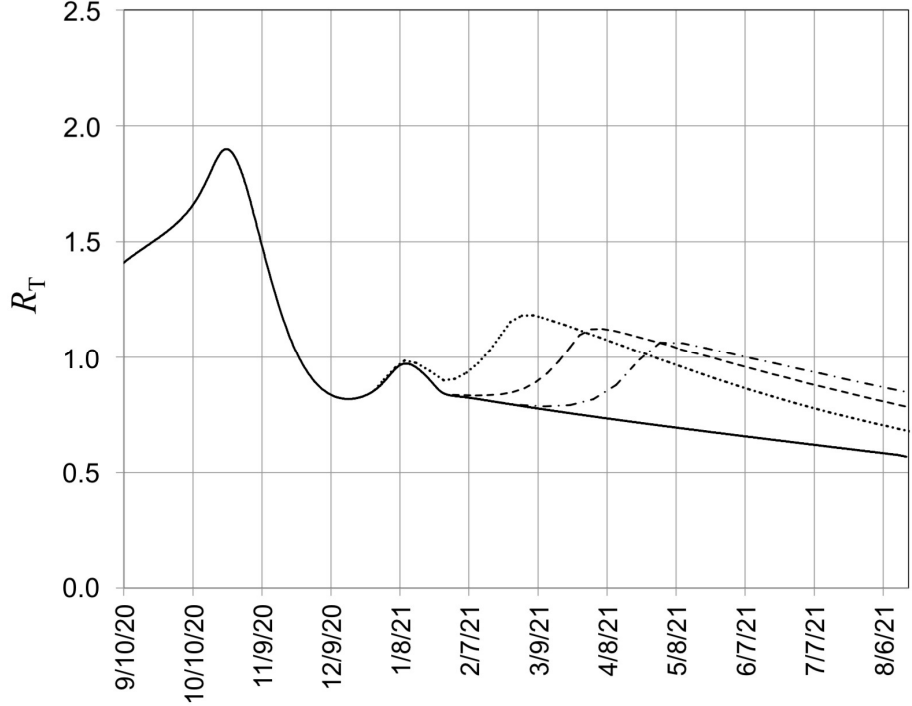

Fig. 9  Reproduction number $R_T$ as a function of time corresponding to the different scenarios of variation of the diffusion coefficient illustrated in Fig. 6.

We finally observe that the integral term in Eq. 16 can be thought as the average number Γ of people, not yet infected ("susceptible people"), that a single positive individual meets throughout his lifetime, since the onset of his infection. Thus, we can conclude that $R_T$ is proportional to Γ, the proportionality constant being $\rho_0 \sigma^2$, i.e. the probability of a single infection event.

Nowadays, the Covid–19 outbreak in the Italian Regions we have analyzed (Lombardia, Sicily, and Lazio) is actually characterized by similar values of $R_T$, all of them being below 1: 0.87 for Lombardia, 0.82 for Sicily, 0.79 for Lazio, 0.83 for Italy as a whole. However, people mobility levels corresponding to these similar $R_T$ values are different Region by Region. This situation is clearly illustrated in Fig. 10, where $R_T$ is plotted as a function of Γ (the number of susceptible people that a single positive meets on average throughout his lifetime) for Italy (continuous line), Lombardia (dashed line), Sicily (dot–dashed line), and Lazio (dotted line). The present ($R_T$, Γ) values, as of January the 31$^{th}$, 2021, are plotted as open circle, open lozenges, open square, and open triangle, respectively. We notice that a same value of $R_T$ corresponds, for the examined Regions, to different average number of susceptible people met by a single person positive for the virus during his lifetime. Setting people mobility to the very same level experienced nowadays on average in Italy (Γ ~ 1300), will produce in Lombardia a significant increase of $R_T$ from 0.87 to 1.75, as a direct consequence of the circumstance that the density of inhabitants of Lombardia is double of that of Italy as a whole.

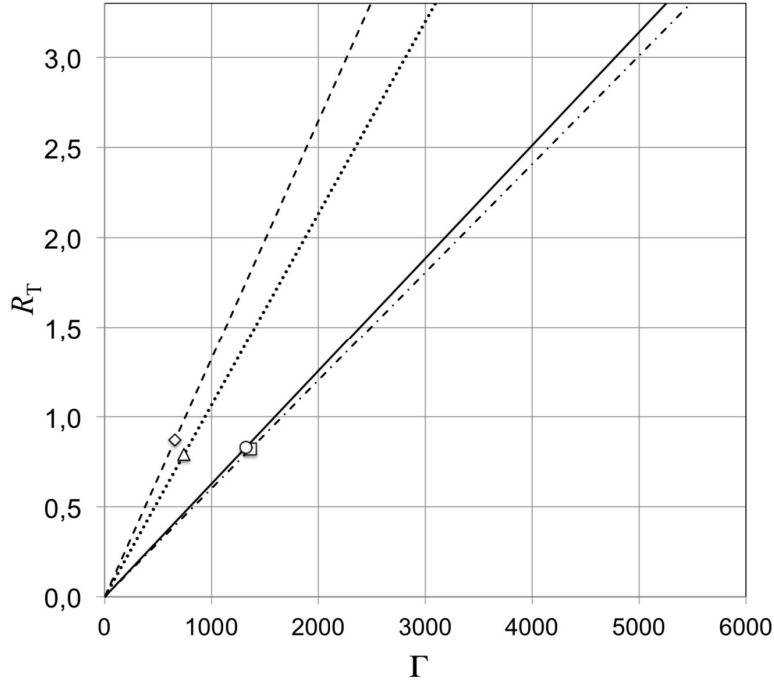

Fig. 10 Reproduction number $R_T$ versus average number $\Gamma$ of susceptible people that a single individual, positive for the virus, meets throughout his lifetime, for Italy (continuous line), Lombardia (dashed line), Sicily (dot–dashed line), and Lazio (dotted line).

*Modelling the effect of virus variants*

Our theoretical description can effectively provide simulation of the influence of virus variants on the time evolution of hospitalized cases and fatalities. A virus variant is expected to be more infective and/or or more severe in terms of the fraction of infected people requiring hospitalization. In the former case we should simply increase the value of the infection cross–section $\sigma^2$, in the latter case it is $f$ that has to be changed. In particular, the effect of virus variants can be simulated by assuming that at, a certain date, the new form of the virus, characterized by different values of $\sigma^2$ and/or $f$, is present in a small fraction of the active positive population. After that time, the concentration $p'$ of people positive for the virus variant will vary with time according to the same set of equations, Eqs. 1 to 6, that are simultaneously used to follow the time variation of the concentration $p$ of people positive for the standard version of the virus. Of course, Eq. 6, has to be modified to take into account for the presence of $p$ and $p'$ in the population of infected people. For virus variant with different transmissibility $\sigma'^2$ and $\sigma^2$, Eq. 6 transforms in:

$$\frac{dc}{dt} = \rho_0 D\sigma^2(\rho_0 - \rho_i - c)p + \rho_0 D\sigma'^2(\rho_0 - \rho_i - c)p' \tag{18}$$

whilst, for virus variant characterized by more severe symptoms but same transmissibility, Eq. 6 becomes:

$$\frac{dc}{dt} = \rho_0 D\sigma^2(\rho_0 - \rho_i - c)(p + p') \tag{19}$$

since the variation $f \to f'$ is included in Eq. 2 describing the time evolution of fraction of $p'$ that requires hospitalization. In all of these approximations, we are assuming that all the other parameters ($q$, $\tau_I$, $\tau_l$, and $\tau_\textrm{l}$) remain unchanged and that vaccine immunization is still effective for either standard or variant form of the virus.

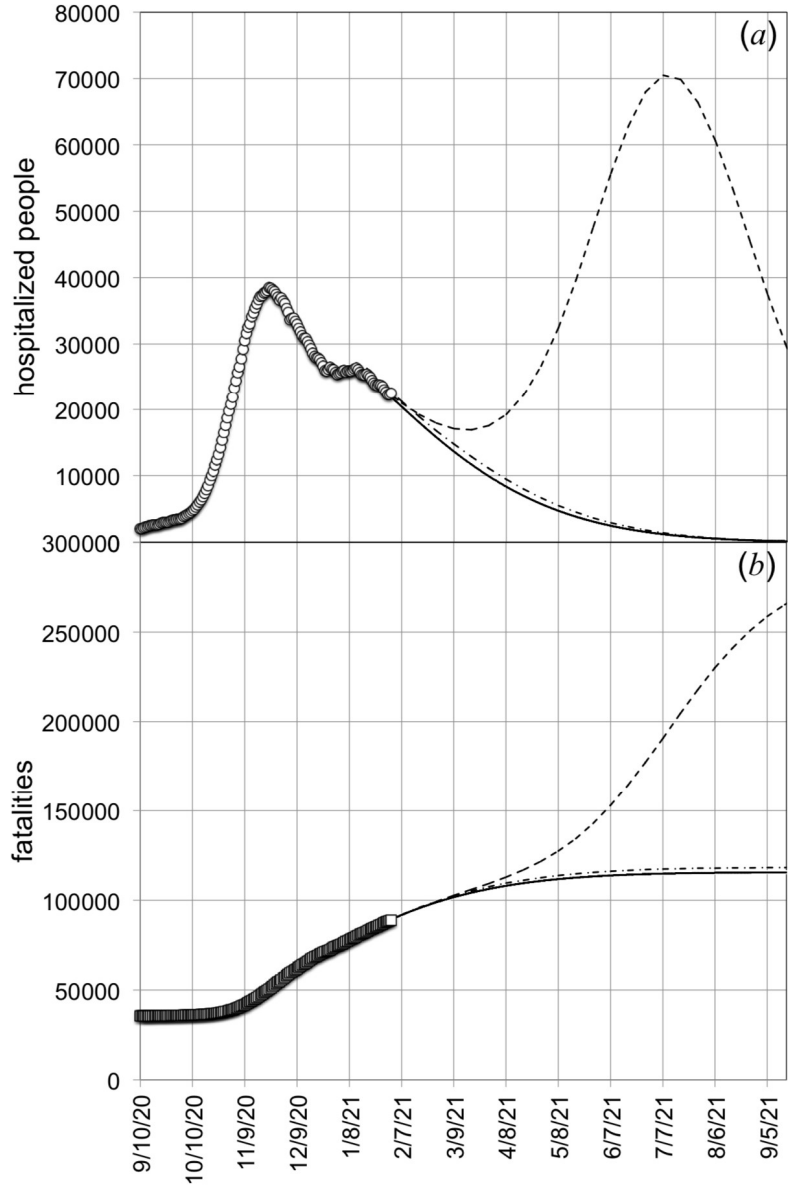

Fig. 11 Simulation of impact of virus variant on the time evolution of hospitalized cases (*a*) and fatalities (*b*) in Italy. Calculations were performed by assuming that 1% of the active positives on January the 15[th], 2021 were affected by a virus variant characterized by an infection cross–section higher by a factor 2 (dashed lines), or by a virus variant producing more severe symptoms, described by an increase by a factor 5 of the fraction of positives requiring hospitalization (dot–dashed lines). Continuous line is the predicted time evolution of cases in the absence of virus variants at the present level of mobility ($D/D_\textrm{L} = 2.1$)

Fig. 11 shows the results of simulating the presence, on January the 15[th], 2021, of people positives to virus variant at a concentration equal to 1% of the total active circulating positives. We performed these calculations by keeping constant the diffusion coefficient to its present value ($D/D_\textrm{L} = 2.1$) and by considering the case of a virus variant with higher transmissibility ($\sigma'^2 = 2\sigma^2$, dashed line in Fig. 11) or producing more serious symptoms ($f' = 5f$, dot–dashed line in Fig. 11). It is evident that the appearance of virus variant having a higher transmissibility is the most dangerous perspective, with respect to the hypothesis that the variant virus characteristics are only limited to the increase of symptoms severity. It is should also emphasized that the increase of cross–section by a factor two is obtained by the increase of the characteristic infection distance, $R$, ($\sigma^2 = \pi R^2$), by just 40%.

## Conclusions

In conclusion, we have shown that the spread of COVID–19 virus can be successfully described by a compartmental model that based on the assumption that the probability of a single infection event is given by the product between the density of inhabitants and a cross–section measuring the distance within which a person positive for the virus can infect a healthy one. Through the model, it is possible to relate the variation of observed hospitalized cases and fatalities to the modification of the mobility restriction measures, by comparing the present behavior to that already experienced during the first wave of the outbreak. The model includes the effect of vaccine immunization and the role of possible virus variants in propagating the infection. The possibility to simulate the time evolution of the observed cases as a function of a diffusion coefficient function is a powerful tool to investigate the best tradeoff between increasing people mobility and effects of vaccination and/or virus variants in order to keep under control the spread of Covid–19 outbreak.

## Data availability

Data and code are available at GitHub (https://github.com/anmio/covid_italy)

## Author contributions

**C.S.:** Conceptualization; Data curation; Formal analysis; Methodology; Resources; Software; Supervision; Visualization; Writing - original draft;
**A.M.M.:** Formal analysis; Visualization; Writing - original draft.

## Additional Information

**Competing Financial Interests**
The authors declare no competing financial interests.

# References

1. Wang, W., Tang, J. & Wei, F. Updated understanding of the outbreak of 2019 novel coronavirus (2019-nCoV) in Wuhan, China. *Journal of Medical Virology* **92**, 441–447 (2020).

2. Wang, H. *et al.* Phase-adjusted estimation of the number of Coronavirus Disease 2019 cases in Wuhan, China. *Cell Discovery* **6**, 1–8 (2020).

3. Tang, B. *et al.* Estimation of the Transmission Risk of the 2019-nCoV and Its Implication for Public Health Interventions. *J Clin Med* **9**, (2020).

4. Tang, B. *et al.* An updated estimation of the risk of transmission of the novel coronavirus (2019-nCov). *Infectious Disease Modelling* **5**, 248–255 (2020).

5. World Health Organization, Coronavirus disease (COVID-2019) situation reports. https://www.who.int/emergencies/diseases/novel-coronavirus-2019/situation-reports/ Accessed on January the 31st, 2021

6. Pullano, G. *et al.* Novel coronavirus (2019-nCoV) early-stage importation risk to Europe, January 2020. *Eurosurveillance* **25**, 2000057 (2020).

7. Chinazzi, M. *et al.* The effect of travel restrictions on the spread of the 2019 novel coronavirus (COVID-19) outbreak. *Science* **368**, 395–400 (2020).

8. Spinella, C. & Mio, A. M. Phenomenological description of spread of Covid-19 in Italy: people mobility as main factor controlling propagation of infection cases. *arXiv:2011.08111 [physics, q-bio]* (2020).

9. Lipsitch, M. *et al.* Transmission Dynamics and Control of Severe Acute Respiratory Syndrome. *Science* **300**, 1966–1970 (2003).

10. Ferguson, N. M. *et al.* Strategies for containing an emerging influenza pandemic in Southeast Asia. *Nature* **437**, 209–214 (2005).

11. Ferguson, N. M. *et al.* Strategies for mitigating an influenza pandemic. *Nature* **442**, 448–452 (2006).

12. Balcan, D. *et al.* Modeling the spatial spread of infectious diseases: The GLobal Epidemic and Mobility computational model. *Journal of Computational Science* **1**, 132–145 (2010).

13. Acioli, P. H. Diffusion as a first model of spread of viral infection. *American Journal of Physics* **88**, 600–604 (2020).

14. Zlatić, V., Barjašić, I., Kadović, A., Štefančić, H. & Gabrielli, A. Bi-stability of SUDR+K model of epidemics and test kits applied to COVID-19. *Nonlinear Dyn* **101**, 1635–1642 (2020).

15. Dipartimento della Protezione Civile: Emergenza Coronavirus https://github.com/pcm-dpc/COVID-19/tree/master/dati-andamento-nazionale Accessed on January the 31st, 2021.

16. Morone, G. *et al.* Incidence and Persistence of Viral Shedding in COVID-19 Post-acute Patients With Negativized Pharyngeal Swab: A Systematic Review. *Front. Med.* **7**, (2020).

17. Ministero della Salute: Piano vaccini anti Covid-19. https://github.com/italia/covid19-opendata-vaccini Accessed on January the 31st, 2021.